\documentclass[journal]{IEEEtran}
\usepackage{url} 
\usepackage{amsmath,amssymb}
\usepackage{booktabs}
\usepackage{adjustbox}
\usepackage{subcaption}
\usepackage{cite}
\usepackage{array}
\usepackage{multirow}

\usepackage{afterpage}
\usepackage{threeparttable}
\usepackage{graphicx}

\newcommand{\Agents}{\ensuremath{A }}
\newcommand{\NumAgents}{\ensuremath{\mathsf{n}}}
\newcommand{\AdversaryAgents}{\ensuremath{A_\mathsf{d}}}
\newcommand{\LegitimateAgents}{\ensuremath{A_\mathsf{l}}}
\newcommand{\PossiblePlans}{\ensuremath{P_\mathsf{a}}}
\newcommand{\Plan}{\ensuremath{p_\mathsf{a,i}}}
\newcommand{\SelectedPlan}{\ensuremath{s_\mathsf{a}}}
\newcommand{\GlobalResponse}{\ensuremath{g}}
\newcommand{\DiscomfortCost}{\ensuremath{D}}
\newcommand{\DiscomfortFunction}{\ensuremath{f_D}}
\newcommand{\InefficiencyCost}{\ensuremath{I}}
\newcommand{\InefficiencyFunction}{\ensuremath{f_I}}

\newcommand{\NumPlans}{\ensuremath{k}}
\newcommand{\PlanSize}{\ensuremath{d}}

\makeatletter
\renewcommand{\section}{\@startsection{section}{1}{\z@}
  {-2ex plus -0.5ex minus -.2ex}
  {1ex plus 0.3ex}
  {\normalfont\normalsize\bfseries}}
\makeatother

\begin{document}

\title{Optimization under Attack: Resilience, Vulnerability, and the Path to Collapse}

\author{
    \IEEEauthorblockN{Amal Aldawsari\textsuperscript{1,2,*}, Evangelos Pournaras\textsuperscript{1}} \\
    \IEEEauthorblockA{\textsuperscript{1}School of Computer Science, University of Leeds, Leeds, UK} \\
    \IEEEauthorblockA{\textsuperscript{2}College of Computer Science and Engineering, University of Hail, Hail, KSA}
    \thanks{\textsuperscript{*}Corresponding author: Amal Aldawsari. \\
    E-mail address: ml21aa2a@leeds.ac.uk; aa.aldosery@uoh.edu.sa.}
}
\maketitle

\begin{abstract}
Optimization is critical for improving the operations of large-scale socio-technical infrastructures such as those found in energy, mobility, and information systems. In particular, understanding the performance of multi-agent discrete-choice combinatorial optimization under distributed adversarial attacks is a compelling and underexplored problem. Multi-agent systems involve a large number of remote control variables that can influence the cost-effectiveness of distributed optimization heuristics. This paper unravels, for the first time, the trajectories of distributed optimization from resilience to vulnerability, and finally to collapse under varying adversarial influence. Using real-world and synthetic data to generate over 112 million multi-agent optimization scenarios, we systematically assess how the number of agents with varying levels of adversarial severity and network positioning influences optimization performance, with particular attention to the impact on Pareto optimality. With this large-scale dataset, made openly available as a benchmark, we disentangle how optimization systems remain resilient to adversaries and which adversary conditions make optimization vulnerable or cause collapse. These findings can support the design of self-healing strategies for fault tolerance and fault correction, addressing a critical gap in adversarial distributed optimization.
\end{abstract}


\section{Introduction} \label{sec1}
The rapid development of Internet of Things (IoT) applications has brought transformative changes in numerous domains, ranging from smart cities to industrial automation and healthcare \cite{perwej2019internet}. These applications involve vast networks of interconnected devices that generate substantial amounts of data, and require efficient and distributed decision-making \cite{ali2018applications}. Distributed optimization is essential in such scenarios, as it enables multiple agents to collaborate effectively without centralized coordination. This approach ensures scalability, reliability, and robust performance across diverse applications, including energy management, autonomous systems, and federated learning \cite{yang2019survey, nedic2009distributed, boyd2011distributed, zhou2021asynchronous, zhou2021graph}.

While distributed optimization is essential for achieving system-wide objectives, most algorithms rely on assumptions of rational, cooperative, and non-adversarial agents contributing toward the global objective without prioritizing their individual goals over collective outcomes. However, real-world scenarios often deviate from these assumptions; adversary agents can disrupt optimization by introducing inaccuracies, manipulating decisions, or compromising functionality \cite{gupta2020fault, ravi2019case, sundaram2018distributed, lu2020distributed}. For example, in smart grid systems where autonomous agents (e.g., households) collaborate to manage energy distribution and prevent power outages, certain households may act adversarially by manipulating consumption data to prioritize their self-interest, i.e., their thermal comfort. Such disruptions distort energy allocation, leading to inefficient resource use and blackouts \cite{ravi2019case, sundaram2018distributed,  rajabi2021resilience, fanitabasi2018review, Pournaras2016}.

Several studies have examined multi-agent optimization in continuous-choice frameworks  \cite{gupta2020fault,ravi2019case, sundaram2018distributed, lu2020distributed, fanitabasi2018review, gonzalez2018multi}. While stochastic optimization techniques address noise, they lack mechanisms to counteract strategic manipulations by adversarial agents \cite{hodgkinson2021multiplicative, spall2012stochastic}. Such manipulations amplify system vulnerability by prioritizing individual goals over collective objectives, which can lead to inefficiencies, instability, and eventual system collapse \cite{hancock2022avoiding, zan2023adversarial}. Despite these risks, adversarial disruptions in discrete-choice settings have received little attention.

This paper studies multi-agent systems in discrete-choice combinatorial optimization under adversarial conditions, with the aim to unravel the trajectories of resilience, vulnerability, and collapse, offering a novel framework to understand system optimization behavior. In this context, resilience refers to the system ability to maintain performance despite adversarial influence; vulnerability captures the emergence of inefficiencies due to intolerable adversarial behaviors; and collapse occurs when the system significantly under-performs as a result of failure to cope with adversarial behaviors. This study evaluates the impact of adversarial agents on optimization performance, analyzing critical parameters such as the number of adversarial agents, their behavioral severity, and their network positions. By systematically examining these dynamics, the paper provides critical insights into the conditions under which systems transition from resilience to collapse, offering new insights for developing self-healing, fault-tolerant, and fault-correcting strategies in adversarial environments \cite{9466159}.

The main contributions of this paper are outlined as follows:
\begin{itemize}
    \item An adversarial model for discrete-choice multi-objective optimization problems.
    \item A novel evaluation framework for characterizing resilience, vulnerability, and collapse in distributed optimization systems under adversarial influence.
    \item A comprehensive evaluation of the adversarial impact on system efficiency and agent discomfort. 
    \item The first open large-scale benchmark datasets for discrete-choice adversarial optimization, generated from over 112 million experiments on real-world and synthetic inputs using the proposed adversarial model.
    \item New insights into how adversarial scale, severity, and structural positioning affect system optimality, including resilience and vulnerability thresholds, Pareto trade-offs, and structural vulnerabilities across diverse scenarios.
    \item An open-source software artifact implementing the proposed adversarial model within the I-EPOS system\footnote{Available at \url{https://github.com/epournaras/EPOS}}, extending its functionality to support heterogeneous agent behaviors.
\end{itemize}

The rest of this paper is organized as follows: Section \ref{sec2} reviews related work on adversarial distributed optimization. Section \ref{sec3} introduces the proposed adversarial model and problem formulation, including the network model and optimization challenge. Section \ref{sec4} outlines the experimental methodology and evaluation metrics. Section \ref{sec5} presents the key findings, analyzing system behavior under varying adversarial conditions. Finally, Section \ref{sec6} concludes the paper, discusses its limitations, and outlines directions for future research.

\section{Related Work} \label{sec2} 
Resilience in distributed multi-objective optimization plays a critical role across domains such as smart grids, transportation, logistics, and communication networks, where robust and adaptive systems are crucial for ensuring operational efficiency \cite{yazdani2023techno, qiao2023multi}. Convex distributed optimization has received significant attention, with a focus on addressing challenges posed by adversary agents, network structures, and varied application domains \cite{yang2019survey, fanitabasi2018review, patari2021distributed}. Earlier work examined the robustness and vulnerability of consensus-based distributed optimization, focusing on addressing limitations related to adversary behavior, network topology, objective functions, and application domains \cite{fanitabasi2018review, fu2021resilient, zhang2023accelerated}. The presence of adversary agents significantly impacts the performance of distributed optimization models. These agents disrupt optimization by slowing convergence, manipulating data, or withholding participation, resulting in suboptimal performance \cite{gupta2020fault, gonzalez2018multi}. Table \ref{t1} provides a comparative analysis of related work on distributed optimization under adversarial conditions. It highlights key aspects such as the type of adversarial behavior\footnote{A malicious node sends the same value to all its neighbors at each time step, whereas a Byzantine node may send different values to different neighbors.}, attack targets, system knowledge\footnote{Knowledge levels: full—complete knowledge of the network and agent objectives; partial—access to limited neighbor information; local—only own state or data is known.}, network structures, and the impact on overall performance.

\begingroup
  \renewcommand{\arraystretch}{1.5}
\begin{table*}[htbp]
  \caption{Comparison of literature on resilience in distributed optimization}
  \centering
  \begin{adjustbox}{width=\textwidth}
    \begin{tabular}{%
      >{\centering\arraybackslash}m{1.8cm}
      >{\raggedright\arraybackslash}m{1.8cm}
      *{15}{>{\raggedright\arraybackslash}p{1.5cm}}
    }
      \toprule
      & & \cite{sundaram2018distributed}
      & \cite{kuwaranancharoen2020byzantine}
      & \cite{su2020byzantine}
      & \cite{lin2020robustness}
      & \cite{figura2021adversarial}
      & \cite{zheng2021vulnerability}
      & \cite{ishii2022overview}
      & \cite{yemini2022resilience}
      & \cite{du2023distributed}
      & \cite{zhao2019resilient}
      &\cite{uribe2019resilient, turan2020resilient}
      & \cite{ravi2019case}
      & \cite{gentz2016data}
      & \cite{ding2018consensus, ding2021differentially}
      & \textbf{This work}\\
      \midrule

     \multirow{4}{=}{\rotatebox{90}{\shortstack{\bfseries Adversarial\\ \bfseries behavior}}}
        &  \textbf{Malicious}
          & \checkmark &  &  &\checkmark &\checkmark &  &  &\checkmark &\checkmark &\checkmark &  &\checkmark &\checkmark &  &\checkmark \\
        &  \textbf{Byzantine}
          &\checkmark &\checkmark &\checkmark &  &  &  &  &  &\checkmark &\checkmark &  &  &  &  &  \\
        &  \textbf{Cyber‐attacks}
          &  &  &  &  &  &\checkmark &\checkmark & \checkmark&  &\checkmark &  &  &  &  &  \\
        &  \textbf{Eavesdropping}
          &  &  &  &  &  &  &  &  &  &  &  &  &\checkmark &  &  \\
      \midrule

      \multirow{4}{3cm}{\rotatebox{90}{\shortstack{\bfseries Adversary\\ \bfseries target}}}
        & \textbf{Consensus}
          &\checkmark &\checkmark &  &  &\checkmark &  &\checkmark &  &  &\checkmark &  &\checkmark &\checkmark &  &  \\
        & \textbf{Information Exchange}
          &  &  &\checkmark &  &  &  &  &\checkmark &  &\checkmark &  &  &\checkmark &  &  \\
        & \textbf{System Objective}
          &  &  &  &  &  &\checkmark &  &\checkmark &  &  &  &  &  &  &\checkmark \\
        & \textbf{Observation}
          &  &  &  &\checkmark &  &  &  &  &  &  &  &  &  &  &  \\
      \midrule

      \multirow{3}{4cm}{\rotatebox{90}{\shortstack{  \bfseries Knowledge\\ \bfseries of the \\ \bfseries system}}}
        & \textbf{Full}
          &\checkmark &\checkmark &\checkmark &  &  &\checkmark &\checkmark &  &  &  &  &  &  &\checkmark \\
        & \textbf{Partial}
          &  &  &  &  &\checkmark &  &\checkmark &  &  &  &  &  &\checkmark &  &  \\
        & \textbf{Local}
          &  &  &  &  &  &  &  &  &  &\checkmark &\checkmark &  &  &\checkmark \\
      \midrule

      \textbf{Directed Network}
        &  
          &  &\checkmark &\checkmark &\checkmark &  &  &  &  &\checkmark &  &\checkmark &\checkmark &  &  &\checkmark \\
      \midrule

      \multirow{4}{3cm}{\rotatebox{90}{\shortstack{ \bfseries Performance\\ \bfseries measure}}}
        & \textbf{Convergence}
          &  &\checkmark &\checkmark &  &  &  &\checkmark &\checkmark &\checkmark &\checkmark &\checkmark &\checkmark &\checkmark &\checkmark &  \\
        & \textbf{Distance to Optimality}
          &\checkmark &\checkmark &\checkmark &  &  &  &  &  &  &  &  &  &  &\checkmark &  \\
        & \textbf{Reward/Utility}
          &  &  &  &\checkmark &\checkmark &\checkmark &  &  &  &\checkmark &  &  &  &  &  \\
        & \textbf{Efficiency}
          &  &  &  &  &  &  &  &  &  &  &  &  &  &  &\checkmark \\
      \midrule

     \textbf{\shortstack{ \\ Algorithms /\\Techniques for\\Optimization}}
        &  
          &Local filtering
          &Distance– based filtering
          &Local filtering
          &  Gradient –based (Deep Q‐learning)
          &  Consensus‐based MARL
          &  None
          &  Mean subsequence reduced
          &  Probabilistic trust– based \&\ projection–based 
          &  Markov switching topology \& Push‐ DIGing
          &  Resilience with trusted agents \& dominating set 
          &  Primal– Dual
          &  FROST
          &  Randomized gossip
          &  Differentially private gradient tracking
          &  I‐EPOS \\
      \bottomrule
    \end{tabular}
  \end{adjustbox}
  \label{t1}
\end{table*}

\endgroup

\subsection{Adversary Agents in Distributed Optimization}
Yang et al. \cite{yang2019survey} provide a comprehensive survey on distributed optimization. Notable advancements include extensions of consensus-based protocols by Sundaram et al. \cite{sundaram2018distributed} and Kuwaranancharoen et al. \cite{kuwaranancharoen2020byzantine}, which address adversarial threats in convex optimization. Su et al. \cite{su2020byzantine} enhance these methods with decentralized architectures and explore adversarial influence on global objectives. However, these approaches assume adversary agents have full knowledge of the network topology and the private functions of all agents. This coordination among adversaries compromises the privacy of the agents in the system.

\subsection{Adversarial Attacks in Multi-Agent Systems}
Adversarial attacks significantly impact reinforcement learning (RL) systems across applications such as robotics, video games, and smart grids, undermining system stability and performance \cite{guesmi2023physical, ali2023survey}. Lin et al. \cite{lin2020robustness} demonstrate how adversarial perturbations affect cooperative multi-agent RL (c-MARL), showing its vulnerability compared to single-agent RL. Figura et al. \cite{figura2021adversarial} highlight how a single adversary can influence consensus-based c-MARL systems, disrupting team objectives. Zheng et al. \cite{zheng2021vulnerability} introduce criticality-based perturbations in deep Q-networks, demonstrating substantial performance degradation due to adversarial attacks. These studies show that an adversary can disrupt system operations and manipulate policies, influencing other agents to adopt behaviors aligned with its objectives.
\subsection{Cyber-Attacks and Resilient Control}
Cyber-attacks, including data injection and denial-of-service (DoS) attacks, pose significant threats to distributed optimization by disrupting system operations and consensus mechanisms \cite{ishii2022overview}. To address these challenges, Yemini et al. \cite{yemini2022resilience} introduce trust-based frameworks that mitigate malicious input, ensuring convergence to global optima. Similarly, Du et al. \cite{du2023distributed} and Zhao et al. \cite{zhao2019resilient} propose models relying on trusted agents to counteract adversarial influence. However, the effectiveness of these can be limited in scenarios with intermittent communication, such as ad hoc or robotic networks.
\subsection{Resource Allocation Challenges Under Adversaries}
In distributed resource allocation, adversarial disruptions are typically mitigated using robust optimization and detection mechanisms. Uribe et al. \cite{uribe2019resilient} and Turan et al. \cite{turan2020resilient} propose primal-dual methods that tolerate Byzantine adversaries by identifying and eliminating malicious inputs, achieving resilience for up to 50\% adversary density. Similarly, Ravi et al. \cite{ravi2019case} develop a detection method that uses agents' data values to identify and isolate potential malicious behavior, imposing an upper limit of 50\% adversaries within the network topology. Gentz et al. \cite{gentz2016data} propose a detection method based on hypothesis-testing for insider attackers in randomized gossip-based sensor networks, leveraging statistical analysis of sensor states to identify malicious agents. While these methods enhance resilience against dispersed adversaries, they assume adversarial influence is evenly distributed and may not generalize to scenarios with concentrated or dynamic adversary placement.
\subsection{Multi-Objective Distributed Optimization}
Recent studies have increasingly adopted Pareto-based multi-objective optimization to evaluate system trade-offs in complex infrastructures. Fettah et al. \cite{fettah2024pareto} introduce a Pareto strategy for optimizing distributed generation in power networks. Zhang et al. \cite{zhang2022multi} formulate a multi-objective operational framework that integrates Pareto analysis to enhance resilience thresholds in distribution networks. Similarly, Boindala and Ostfeld \cite{boindala2022robust} propose an optimization approach to balance reliability, cost, and failure risk using Pareto fronts. While these studies underscore the value of Pareto analysis for resilient optimization, they focus on centralized infrastructures and do not address adversarial influence or the complexities of distributed, multi-agent decision-making explored in this work.

\subsection{Privacy-Preserving Distributed Optimization}
Privacy-preserving distributed optimization safeguards sensitive agent information against eavesdropping adversaries using techniques such as differential privacy \cite{Asikis2020,ding2018consensus, ding2021differentially}, homomorphic cryptography \cite{zhang2018enabling, lu2018privacy}, and gradient perturbation \cite{mao2020privacy, chen2023differentially}, to ensure secure information exchange. However, these studies focus on privacy protection rather than adversarial behavior in optimization contexts.
 
\subsection{Combinatorial Optimization Algorithms}
In distributed combinatorial optimization, Hinrichs et al. \cite{hinrichs2014cohda, hinrichs2017distributed} propose COHDA, a combinatorial optimization heuristic designed for multi-agent systems. However, COHDA encounters scalability challenges due to increasing communication overhead as the number of agents grows. The collective learning approach of Pournaras et al. \cite{pournaras2017self} address this with EPOS (Economic Planning and Optimized Selections), a distributed optimization method that enables agents to collaboratively optimize global resource allocation, particularly in participatory sharing economies. Although EPOS ensures privacy, autonomy, and scalability, it faces computational limitations when applied to wide tree structures with multiple child nodes \cite{pournaras2018decentralized}. I-EPOS is the iterative extension of EPOS; it incorporates decentralized iterative back-propagation and localized decision-making to enhance scalability and support plan coordination across deeper and broader network hierarchies \cite{pournaras2020collective, pournaras2018decentralized}.

While COHDA, EPOS, and I-EPOS represent foundational combinatorial optimization approaches, their  system performance under adversarial conditions has not been studied before, despite some limited work on measuring the impact of arbitrary structural faults, which does not focus on agents' behavior~\cite{pournaras2020holarchic}. In this work, we introduce a generic adversarial model applicable to such settings. The model enables a structured evaluation of resilience, vulnerability, and collapse dynamics, and, to the best of our knowledge, is the first to systematically explore adversarial behaviors in discrete-choice combinatorial optimization.

A large body of research to date focuses on continuous distributed optimization, often assuming limited adversary presence or relying on complete graph topologies \cite{gupta2020fault,ravi2019case, sundaram2018distributed, fanitabasi2018review}. Such assumptions do not fully capture the complexity of real-world systems, where distributed structures, heterogeneous agent behaviors, and dynamic adversarial threats are prevalent \cite{zheng2021vulnerability, sundaram2018distributed}. Although existing solutions offer valuable mitigation strategies \cite{uribe2019resilient, du2023distributed}, the lack of comprehensive analyses on inherent system vulnerability, resilience thresholds, and pathways to optimization collapse remains a gap.

To fill this gap, we propose a generic adversarial model to systematically analyze how adversarial agents influence system performance and stability in distributed multi-objective optimization.Our objective is to evaluate how adversarial behavior influences resilience, vulnerability, and collapse, while informing the development of self-healing strategies for robust optimization. Moreover, his work releases the first large-scale, open benchmark dataset designed for evaluating adversarial impacts under discrete-choice optimization settings. To clarify the novelty of our contribution, Table \ref{novelty_comparison} summarizes a comparative analysis with existing work, highlighting key aspects such as discrete decision-making, multi-objective formulation, resilience thresholds, structural vulnerability, and Pareto-based evaluation.

\begin{table*}[ht]
\caption{Comparison of the novelty aspects of this work with related distributed optimization approaches}
  \centering
  \footnotesize
  \begin{adjustbox}{width=\textwidth}
    \begin{tabular}{{lcccccccccccccccc}}
      \toprule
      \textbf{Criteria} & \textbf{This work} & \textbf{\cite{sundaram2018distributed}} & \textbf{\cite{kuwaranancharoen2020byzantine}} & \textbf{\cite{su2020byzantine}} & \textbf{\cite{lin2020robustness}} & \textbf{\cite{figura2021adversarial}} & \textbf{\cite{zheng2021vulnerability}} & \textbf{\cite{ishii2022overview}} & \textbf{\cite{yemini2022resilience, du2023distributed, zhao2019resilient}} & \textbf{\cite{uribe2019resilient, turan2020resilient}} & \textbf{\cite{ravi2019case}} & \textbf{\cite{gentz2016data}} & \textbf{\cite{ding2018consensus, ding2021differentially}} & \textbf{\cite{fettah2024pareto, zhang2022multi,boindala2022robust}} & \\
      \midrule
      Multi-objective optimization & \checkmark & X & X & \checkmark & X & \checkmark & X & X & \checkmark & \checkmark & \checkmark & X & \checkmark   & \checkmark \\
        \midrule
      Discrete decision-making & \checkmark & X  & X & X& \checkmark & \checkmark  & \checkmark & X& X  & X & X & X  & X & X\\
       \midrule
      Resilience \& vulnerability thresholds & \checkmark & \checkmark & \checkmark & \checkmark  & X  & X  & X & \checkmark & \checkmark & \checkmark & \checkmark & \checkmark  & X  & \checkmark  \\
       \midrule
      Structure analysis & \checkmark & \checkmark & X  & \checkmark & X & X & X & \checkmark& \checkmark  & X & X & X  & X &X \\
       \midrule
      Pareto analysis & \checkmark & X &  X & X & X & X & X & X & X & X & X & X  & X  & \checkmark\\
      \bottomrule
    \end{tabular}
  \end{adjustbox}
  \label{novelty_comparison}
\end{table*}

\section {Adversarial Distributed Optimization} \label{sec3}
\subsection{Problem Formulation}
Resilience in distributed optimization is essential for maintaining system performance under adversarial conditions. Adversary agents disrupt operations, degrade efficiency, and increase vulnerability. This raises key questions: How do adversary agents influence the efficiency and stability of distributed optimization systems? What thresholds of adversarial behavior lead to transitions from resilience to vulnerability or collapse? How do parameters such as adversary density, adversarial severity, and network positioning influence these transitions?

To address these challenges, we propose a generic adversarial distributed optimization model tailored to discrete-choice scenarios. Our model investigates the trade-offs between system-wide and individual agent goals in adversarial settings. It incorporates key parameters, including the scale of adversaries, behavioral severity, and structural positioning, providing a robust framework to evaluate system vulnerability and resilience. Through this study, we identify critical thresholds for transitions from resilience to collapse, providing an in-depth understanding of system behavior under adversarial influence. These findings inform the development of self-healing strategies that enhance fault-tolerance and mitigate adversarial impacts across diverse distributed optimization applications.

\subsection{Network Model}
Consider a network with \NumAgents{} agents, denoted by \Agents, each identified by a unique ID in the set $\{1,2,3,...,\NumAgents\}$. The network topology is represented as a connected graph $G = (\Agents, E)$, where \Agents{} is the set of agents, and $E$ is the set of edges, with $(j, i) \in E$ not necessarily implying $(i, j) \in E$. Agents interact within a self-organized network through bidirectional
communication to exchange information and update their states to align with system goals.

Table \ref{t2} summarizes the notations used throughout the paper to formalize the network model and optimization framework.

\begin{table}[ht]
 \caption{Nomenclature utilized in this research}
    \label{t2}
     \centering
    \begin{tabular}{ll}
    \toprule
    \textbf{Notation} & \textbf{Description} \\
    \midrule
    \Agents  & set of agents in the network \\
    \NumAgents$ = |\Agents|$  & number of agents  \\
    $\AdversaryAgents \subseteq \Agents$ & set of adversary agents  \\
    $\LegitimateAgents \subseteq \Agents$  & set of legitimate agents \\
    \PossiblePlans & set of possible plans of agent $a$  \\
    \Plan$ \in \PossiblePlans$ & plan $i$ of agent $a$ \\
    \NumPlans & number of plans \\
    \PlanSize & size of plan \\ 
    \SelectedPlan & selected plan of agent $a$ \\
    \GlobalResponse & global response \\
    \DiscomfortCost & discomfort cost \\
    \DiscomfortFunction & discomfort cost function \\
    \InefficiencyCost & inefficiency cost \\
    \InefficiencyFunction & inefficiency cost function\\
    \bottomrule
    \end{tabular}
\end{table}

\subsection{Optimization Framework in Discrete-Choice Scenarios}
In distributed discrete-choice optimization, each agent \(a \in \Agents\) selects one option from a finite set of \NumPlans{} alternatives referred to as \textit{possible plans} \PossiblePlans \(\subset \mathbb{R}^d\). Each plan \Plan \(\in \PossiblePlans\) is a sequence of size \PlanSize{} that represents a decision configuration, such as resource allocation or scheduling. These plans reflect the agent's potential future operations, from which the agent selects one, denoted as \SelectedPlan. The collective outcome is captured by the \textit{global response} \(\GlobalResponse = \sum_{a \in \Agents} \SelectedPlan\), which aggregates all selected plans of all agents to evaluate system-level performance. For instance, in power grid systems, each household acts as an agent with multiple plans representing alternative appliance energy consumption levels~\cite{Fanitabasi2020}. Each household selects one plan, contributing to the total energy consumption, which represents the global response (\GlobalResponse) in power grid systems.

Agents aim to balance their individual preferences with system-wide goals, which often involve conflicting criteria. Each agent $a$ has an individual preference for its plans, quantified by the \textit{discomfort cost} (\DiscomfortCost), such that \DiscomfortCost\(_{a,i} = \DiscomfortFunction(\Plan)\), where \DiscomfortFunction{} measures how undesirable plan \ensuremath{i} is for agent \ensuremath{a} based on the agent’s preferences; lower costs indicate more preferred plans. Each agent $a$ evaluates the costs \DiscomfortCost{} for each possible plan \Plan $\in$ \PossiblePlans{} and the plan with the minimum cost is selected. 

While agents aim to minimize their own discomfort, they may also consider system-wide metrics such as the \textit{inefficiency cost} (\InefficiencyCost). The inefficiency cost (\InefficiencyCost) is the measure used to evaluate the collective system-wide performance based on the aggregated responses of all agents. It represents the system-wide performance inefficiency that agents aim to minimize through coordinated decision-making: $\InefficiencyCost = \InefficiencyFunction\left(\sum_{a=1}^{\NumAgents}(\SelectedPlan)\right)$. Each agent selects a plan that minimizes a weighted combination of individual discomfort and system-wide inefficiency, as shown in Equation \ref{eq1}.

\begin{align}
    \SelectedPlan &= \arg\min_{i=1}^{\NumPlans} \left( \alpha_a \cdot \InefficiencyCost_{a,i} + \beta_a \cdot \DiscomfortCost_{a,i} \right) \nonumber \\
    &= \arg\min_{i=1}^{\NumPlans} \left[ 
    \alpha_a \cdot \InefficiencyFunction\left(s_1 + s_2 + \cdots + s_{\NumAgents}\right) \right. \nonumber\\
    &\quad \left. + \text{ }\beta_a \cdot \DiscomfortFunction\left(\DiscomfortCost_{1,s}, \DiscomfortCost_{2,s}, \ldots, \DiscomfortCost_{\NumAgents,s}\right) \right] \text{,}\label{eq1}
\end{align}

\begin{align*}
\text{where} \quad  \alpha_a + \beta_a &= 1 \quad \text{\&} \quad \alpha_a, \beta_a \in [0, 1]
\end{align*}

The behavior of agent $a$ is modeled by the corresponding weights $\alpha_a$ and $\beta_a$, which represent agent’s priorities between minimizing system-wide inefficiency and personal discomfort, respectively. A higher weight indicates a higher preference for minimizing the corresponding objective. On the other hand, a weight of 0 means that the corresponding objective is not considered. For instance, an agent with \(\alpha_a = 1\) and \(\beta_a = 0\) behaves altruistically, prioritizing global goals. Conversely, \(\alpha_a = 0\) and \(\beta_a = 1\) define a selfish agent focusing solely on its individual preference.

\subsection {Adversarial Distributed Optimization Model} 
We propose an adversarial model applicable across a range of combinatorial optimization scenarios. In this model, the agent population \Agents{} is partitioned into two disjoint subsets: legitimate agents $\LegitimateAgents$ and adversary agents $\AdversaryAgents$, such that $\Agents = \LegitimateAgents \cup \AdversaryAgents$. While legitimate agents, $\LegitimateAgents \subseteq \Agents$, align their actions with system-wide objectives to optimize overall performance, adversary agents, $\AdversaryAgents \subseteq \Agents$, prioritize their individual interests over collective system goals by adapting their behavior to maximize personal benefits. For instance, in a bike-sharing system~\cite{pournaras2018decentralized}, optimization ensures a balanced distribution of bikes across stations to meet user demand. Legitimate users may select pick-up and drop-off stations while considering system-wide efficiency, maintaining network equilibrium. In contrast, adversary users prioritize their own convenience, selecting stations solely based on personal preference, leading to imbalances such as empty or overloaded stations, and ultimately degrading overall efficiency and user satisfaction.

Adversarial behavior is modeled by adjusting the weight $\beta_a$ in the agent’s decision function (Equation \ref{eq1}). Legitimate agents are assigned $\beta_l = 0$, fully aligning with system goals, while adversary agents are assigned $\beta_d > 0$, increasing emphasis on personal discomfort minimization at the expense of system-wide efficiency. This parameterization allows adversarial behavior to be modeled in a continuous space, from fully altruistic to fully selfish.

By varying the distribution and severity of adversarial weights across the agent population, our model enables systematic analysis of resilience, vulnerability, and collapse in distributed optimization. This includes assessing how the impact of adversarial behavior is linked to the network positioning of the adversary agents and legitimate agents. Adversarial behavior amplifies the inefficiency cost \InefficiencyCost{}, reflecting the trade-off between minimizing individual discomfort (\DiscomfortCost{}) and optimizing overall system efficiency (\InefficiencyCost{}). While \DiscomfortCost{} focuses on individual preferences, \InefficiencyCost{} addresses the system-wide inefficiency caused by deviations from optimal resource allocation, underscoring the conflict between individual and collective optimization goals.

\section{Experimental Methodology} \label{sec4}

This section illustrates the distributed optimization method employed as a case study, the experimental setup, the application scenarios, the measured variables and the evaluation metrics. 

This paper proposes a novel adversarial distributed optimization model to study the performance of multi-agent discrete-choice combinatorial optimization
under distributed adversary attacks.

\subsection{Distributed Optimization Method}\label{sec4-1}
The adversarial distributed optimization model is implemented within the \textit{Iterative Economic Planning and Optimized Selections} (I-EPOS) framework. I-EPOS is a discrete-choice distributed combinatorial optimization algorithm for large-scale multi-agent networks \cite{pournaras2018decentralized, pournaras2020collective}. It employs a self-organized, multi-level hierarchical structure to enable efficient communication, coordination, and scalability while minimizing communication overhead \cite{pournaras2013multi}.

I-EPOS enables the agents to iteratively coordinate their choices in collective decision-making. Each iteration consists of two distinct phases: a bottom-up phase and a top-down phase. During the bottom-up phase, agents select plans based on the aggregated choices of agents in the branch beneath, as well as the selections made by all agents in the previous iteration. Conversely, the top-down phase addresses incomplete knowledge from higher branches in the hierarchy, enabling agents to revert to previous selections if no cost reduction is achieved. This process continues until a predefined iteration limit is reached or no further improvement in the optimization objective occurs \cite{pournaras2018decentralized, pournaras2020collective}. This iterative coordination mechanism addresses the inherent complexity of multi-agent optimization, particularly under non-linear cost functions and incomplete knowledge of other agents’ choices. These conditions make the optimization problem NP-hard \cite{pournaras2018decentralized}, requiring distributed coordination methods that allow agents to refine their decisions based on both local preferences and system-wide impact.

The hierarchical network is structured as an acyclic graph, where each parent agent aggregates responses from its children by avoiding double counting. This design ensures efficient coordination when optimizing individual decisions and system-wide objectives \cite{pournaras2018decentralized, pournaras2013multi, pournaras2020holarchic}. I-EPOS is well-suited for adversarial scenarios in large-scale distributed systems due to its scalability, adaptability to diverse agent behaviors, and potential mitigate adversarial conditions \cite{pournaras2020collective, pournaras2024collective}.

To apply the proposed adversarial model, the I-EPOS framework was extended to support heterogeneous agent behaviors. The original implementation assumed uniform agent preferences across the population. We enhanced the framework to allow agents to configure individual decision-making weights. This enhancement enables modeling adversarial agents with varying levels of behavioral severity and is made available as an open-source artifact to support reproducibility and future research\footnote{Available at \url{https://github.com/epournaras/EPOS}.}.

\subsection{Experimental Setup}
Experiments are conducted using multiple HPC servers with varying configurations that support large-scale experimentation and ensure computational efficiency. These include high-memory nodes (up to 768 GB) and multi-core processors (up to 40 cores per node). In addition to these servers, the University of Leeds ARC4 system\footnote{ARC4 is an HPC cluster at Leeds providing a Linux-based HPC service based on CentOS 7. More information: \url{https://arcdocs.leeds.ac.uk/systems/arc4.html}} is utilized. The ARC4 system includes two nodes, each equipped with 40 cores, 768 GB of memory, and 800 GB of storage, providing robust computational capacity for large-scale experiments.

\subsection{Application Scenarios}

Adversarial distributed optimization is studied in three application scenarios based on real-world data and a synethic dataset. Table~\ref{tab:datasets} provides an overview of the datasets used in this research, including the agent populations, the number and size of plans per agent, agent representation and the interpretation of discomfort and inefficiency costs within each application domain. These costs are defined explicitly through the optimization objectives and reflect domain-specific constraints. Further mathematical definitions of cost functions are provided in \ref{appendix:cost-functions}.

\begin{table*}[htbp]
  \centering
   \footnotesize
  \caption{Description of the datasets and experimental setup in this research}
  \begin{adjustbox}{width=\textwidth}
  \begin{tabular}{p{2cm}p{1cm}p{1cm}p{1cm}p{1.5cm}p{3.3cm}p{2.7cm}p{2.2cm}} 
    \toprule
    \textbf{Dataset Name} & \textbf{No. Agents} & \textbf{No. Plans} &  \textbf{Plan Size} & \textbf{Agents}&\textbf{Discomfort Cost}& \textbf{Inefficiency Cost} & \textbf{Total Experiments} \\
    \midrule
    Energy & 1000 & 10 & 144 & Households & Time shift from intrinsic preference & Variance of energy demand & 3,118,560 \\ 
     \midrule
     Privacy & 72  & 3  & 64 & Smart phone users  & Privacy loss & Mismatch between shared and desired data & 498,780 (2 target signals) \\ 
     \midrule
     Voting  & 266  & 31 & 5  & Voters  & Compromise distance from intrinsic voting preferences &  Polarization & 103,456,800 (120 target signals) \\ 
    \midrule
    Gaussian (synthetic) & 10--100 & 2--10 & 2 & Simulated agents & Ranking distance & Variance & 4,125,000 \\
    \bottomrule
  \end{tabular}
\end{adjustbox}
\label{tab:datasets}
\end{table*}

\subsubsection{Energy-demand dataset}
The energy application scenario uses a dataset derived from simulated zonal power transmission in the Pacific Northwest\footnote{Available upon request at \url{http://www.pnwsmartgrid.org/participants.asp}}. The dataset includes power consumption profiles for 1,000 users, with each user represented by an agent containing 10 possible plans. Each plan comprises a 144-length sequence representing electricity consumption at 5-minute intervals over a 12-hour period. These plans are generated using load-shifting strategies to balance grid load during peak and off-peak hours, reducing strain on the energy system. Plans are ranked by preference scores ranging from 0 to 1, with higher scores reflecting greater alignment with the user's original consumption patterns. 
The inefficiency cost is measured as the deviation in aggregated energy consumption from the desired load-balancing levels, capturing the system ability to maintain stability and efficiency. Adversarial households disrupt the system by selecting plans that counteract load balancing, thereby increasing the risk of grid instability during peak periods.

\subsubsection{Privacy dataset}
The privacy dataset originates from a living-lab experiment at the Decision Science Laboratory\footnote{\url{https://www.descil.ethz.ch}} of ETH Zurich, involving 72 participants evaluating 64 data-sharing scenarios that involve 4 sensor types, data collectors, and contexts \cite{pournaras2024collective}. The data-sharing choices of each participant in the experiment determine three data-sharing plans, representing their intrinsic motivation to share and two rewarded scenarios. The plans are assessed using privacy valuation schemes assigning normalized costs ranging from 0 to 1, where lower costs indicate less privacy compromise. The dataset facilitates testing under high and low privacy-preservation target signals. The inefficiency cost is calculated by the residual sum of squares between the shared and the desired data that measures their mismatch and is an indicator of quality of service supported by the collected data. Adversarial participants disrupt coordination by focusing solely on minimizing their data sharing, under-mining the quality of service of data collectors.

\subsubsection{Voting dataset}
This new dataset is derived from voting data in a regional election with five candidates and 266 voters\footnote{The UK Labour Party Leadership Vote Available at \url{https://preflib.simonrey.fr/datasets}} \cite{majumdar2024score}. Each voter has 31 alternative voting plans, representing ranked preferences among the five candidates.  The optimization focuses on minimizing polarization in the voting outcomes, which refers to reaching the same voting outcome but with compromises that reduce polarization. Polarization here is the inefficiency cost and it refers to the mismatch from a liner ranking of the alternatives in the voting outcomes, although other polarization models could be studied as well~\cite{navarrete2024understanding}. The rationale of linearity is to deviate from concentrating the voters' preferences to two opposing poles. To control for the same voting outcome, 120 target signals are generated from all combinations of values 0, 0.25, 0.5, 0.75, and 1, representing the linear ranking of alternatives. Adversarial behavior occurs when voters prioritize their intrinsic preferences, i.e., no compromises to reduce polarization.

\subsubsection{Synthetic Gaussian dataset}
The synthetic dataset is constructed to evaluate system scalability under controlled, domain-agnostic conditions. It includes 100 agents with 10 generated plans, where each plan is a 100-dimensional vector sampled from a Gaussian distribution $\mathcal{N}(0,1)$. Plans are sorted by their index, with lower indices arbitrary indicating higher agent preference. Discomfort cost is defined as the rank of the selected plan—i.e., a higher index reflects greater deviation from the most preferred option. Inefficiency cost is measured as the variance of the aggregated global response, capturing system-wide imbalance. This synthetic setup allows systematic analysis of adversarial effects across varying agent populations, number/size of plans, and attack configurations.

\subsection{Varying Dimensions and Performed Experiments}
The following dimensions are studied in the performed experiments:
\begin{itemize}
\item \textbf{Scale of adversaries ($|\AdversaryAgents|$):} Incrementally increase the number of adversary agents \AdversaryAgents{} from 1 to \NumAgents{} across all datasets to analyze performance under varying adversary densities; i.e., $\AdversaryAgents=\{ a \mid a \in \Agents\}$.

\item \textbf{Adversarial severity ($\beta$):} The adversarial preference to minimize discomfort cost (\(\beta\)) is varied across 30 levels, with $\beta$ incrementing from 0 to 1 such that $\beta = \frac{b}{30}$ for $b = \{1, 2, 3, ..., 30\}$. 

\item \textbf{Adversary position:} The influence of adversary positions within the hierarchical network is analyzed using two approaches: layer-wise and cumulative structural analysis. A binary tree structure is employed, with each hierarchical layer containing approximately $\log_2 |A|$ agents, where $|A|$ is the total number of agents. The structural analysis evaluates inefficiency costs under varying adversary scales (25\%, 50\%, 75\%, 100\%) at each layer of the hierarchy. The cumulative analysis examines the aggregated impact of adversary agents positioned incrementally in top-down (root-to-leaf) and bottom-up (leaf-to-root) configurations.
\end{itemize}

For each dataset, experiments are conducted across 30 adversarial severity levels (\(\beta\)) with 100 simulation runs per configuration. 

Layer-wise structural analysis is performed at four adversarial proportions \( p \in \{25\%, 50\%, 75\%, 100\%\} \) within each layer of the hierarchical topology, where the number of layers is defined as \( \lceil \log_2 |A| \rceil \). For each layer \( L \), the number of adversarial agents is calculated as \( k_p = \max\left(1,\ \left\lceil \frac{p}{100} \cdot |A_L| \right\rceil \right) \), where \( |A_L| \) denotes the number of agents in layer \( L \). Adversarial configurations are sampled up to 100 combinations per setting to ensure computational feasibility\footnote{The equation assumes up to 100 combinations per layer-percentage, but the binary hierarchy often yields fewer due to limited agents per layer.}. In addition, two cumulative structural attack scenarios are simulated, top-down (root-to-leaf) and bottom-up (leaf-to-root), introducing \( 2 \times |A| \) further experiments per dataset. The total number of experiments per dataset is calculated as:

$\text{Total Experiments} = (30 \times \text{number of signals}) \times \left[(100 \times |A|) + \sum_{L} \sum_{p} \min\left(100,\ \binom{|A_L|}{k_p}\right) + (2 \times |A|)\right]$

For the synthetic Gaussian dataset, experiments vary both agent populations and the number of plans per agent (from 2 to 10). The total number of experiments is computed as: $\text{Total Experiments}_{\text{Gaussian}} = \sum_{i=1}^{10} \left( 10i \times 30 \times 5 \times 50 \right),$ where \( i \) is the number of agents, 30 is the number of severity levels, 5 is the number of plans, and 50 is the number of random structural permutations (i.e., reordering of agents in the tree). 

\subsection{Evaluation Metrics}
\textbf{Optimization objectives:} 
System performance is assessed using three metrics: inefficiency cost \InefficiencyCost{}, discomfort cost \DiscomfortCost{}, and the compromise cost of legitimate agents. Inefficiency cost captures the deviation from optimal system performance. Discomfort cost reflects individual agents' dissatisfaction, i.e. to what extent a plan is not the most preferred one. The compromise cost quantifies the increase in discomfort experienced by legitimate agents due to adversarial influence. It is calculated as the difference in discomfort between scenarios with and without adversaries, highlighting the collective burden legitimate agents bear to maintain system performance.

\textbf{Pareto optimality:}
In multi-objective optimization, the Pareto front defines solutions where no objective can improve without compromising another objective. The knee point on this front identifies the most balanced trade-off between competing objectives. This study uses the Minimum Manhattan Distance (MMD) method to locate the knee point, measuring the distance from each Pareto solution to an ideal reference point where both objectives are optimized. The solution with the smallest distance is selected. This approach ensures a balanced trade-off between discomfort (\DiscomfortCost{}) and inefficiency (\InefficiencyCost{}) in line with established approaches \cite{sun2024knee, li2020knee}.

\textbf{Resilience, vulnerability, and collapse framework:}
To classify system states, the multi-Otsu thresholding method is applied to segment inefficiency and discomfort values into three distinct regions: resilience (low inefficiency), vulnerability (moderate inefficiency), and collapse (high inefficiency). This technique minimizes intra-class variance, offering a robust framework to detect transitions in system performance under adversarial conditions \cite{otsu1979, merzban2019efficient}.

\begin{figure*}[t]
  \centering
  \begin{subfigure}{\textwidth}
    \centering
    \includegraphics[width=0.95\textwidth]{Inefficinecy_knee.pdf}
    \caption{Inefficiency costs}
    \label{fig:inefficinecy}
  \end{subfigure}
  \begin{subfigure}{\textwidth}
    \centering
    \includegraphics[width=0.95\textwidth]{Discomfort_knee.pdf}
    \caption{Discomfort costs}
    \label{fig:discomfort}
  \end{subfigure}
  \begin{subfigure}{\textwidth}
    \centering
    \includegraphics[width=0.95\textwidth]{Compromised_knee.pdf}
    \caption{Compromised discomfort costs for legitimate agents}
    \label{fig:compromised}
  \end{subfigure}
    \caption{Inefficiency, discomfort, and compromised discomfort over adversary scales and severity in the energy, voting, and privacy datasets, , including Pareto knee points and resilience (R), vulnerability (V), and collapse (C) thresholds.}
  \label{fig:combined_plots}
\end{figure*}

\section{Results Analysis and Discussion} \label{sec5}
This section presents the results of extensive experiments evaluating the impact of adversarial agents on multi-objective distributed optimization across the real‐world (energy, voting, and privacy) and synthetic datasets.

\begin{figure*}[htbp]
  \centering
  \begin{subfigure}{0.98\textwidth} 
    \includegraphics[width=\textwidth]{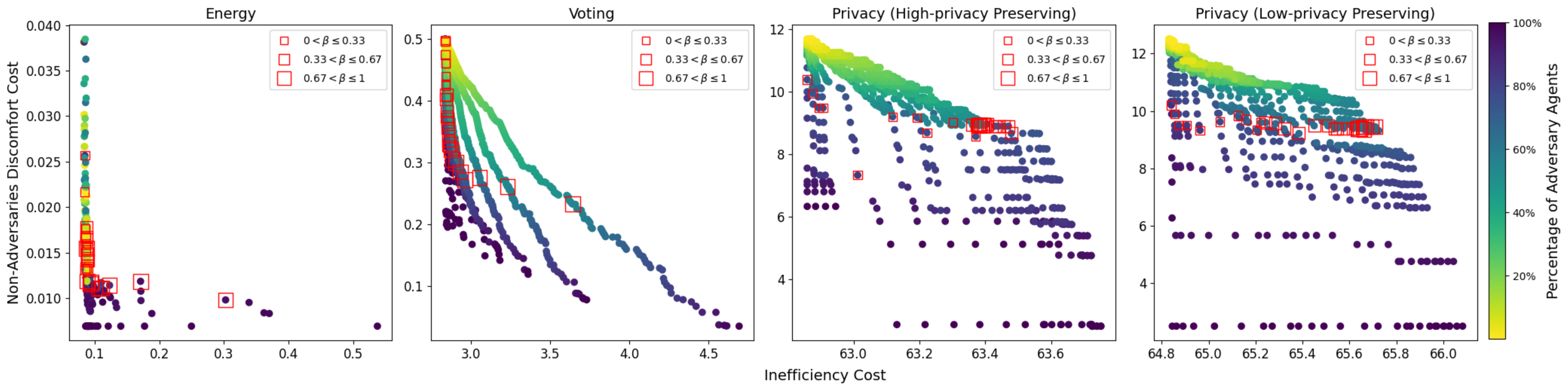}
    \centering
    \caption{Pareto front and knee points across adversarial severity.}
    \label{fig:pareto_level}
  \end{subfigure}

  \begin{subfigure}{0.98\textwidth}
    \includegraphics[width=\textwidth]{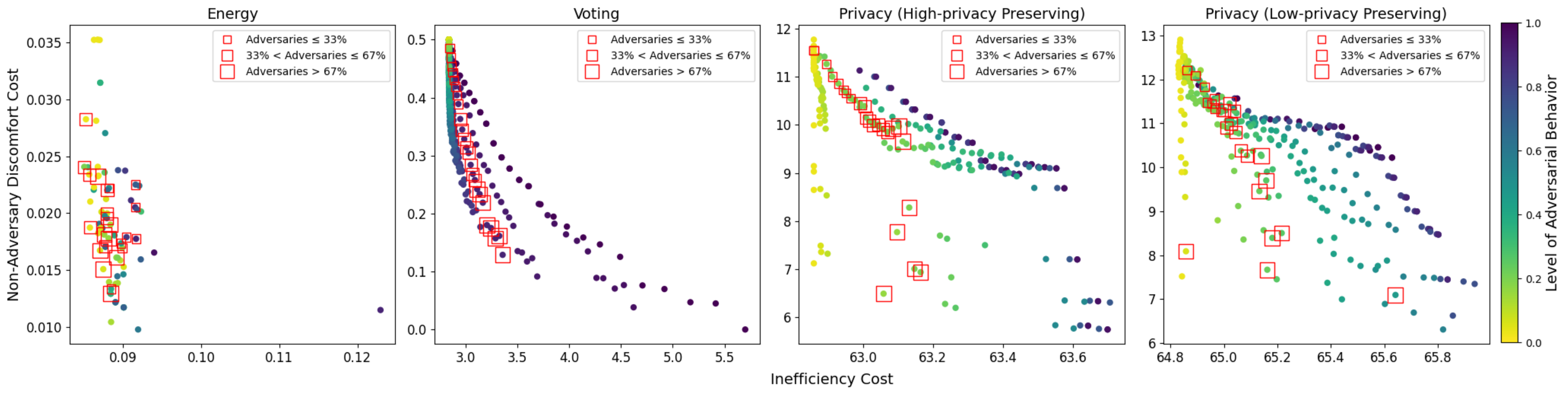}
    \centering
    \caption{Pareto front and knee points across scales of adversaries.}
    \label{fig:pareto_adv}
  \end{subfigure}

\caption{The Pareto Optimality of The energy voting, and privacy datasets}
\label{fig:pareto}
\end{figure*}

\subsection{Resilience Analysis}
Figure \ref{fig:combined_plots} presents inefficiency, discomfort, and compromised discomfort costs across varying adversary scales and severity levels ($\beta$). These metrics capture degradation in system performance, agent satisfaction, and the impact on legitimate agents. The graphs include resilience, vulnerability, and collapse thresholds, with overlaid Pareto knee points. For the inefficiency and compromised‐discomfort plots, knees mark the best trade‐off between those two metrics; on the discomfort heatmaps, the knee indicates the minimal total discomfort for a given inefficiency.

Inefficiency cost (Figure \ref{fig:inefficinecy}) remains low in the resilience zone, especially when adversary ratios are below 30–50\% and $\beta < 0.8$. As adversary scale and severity increase, systems shift from Resilience to Vulnerability and Collapse, with thresholds varying across datasets. Energy and voting maintain ~90\% resilience and only 3–5\% collapse, while privacy configurations show earlier collapse near 20\%. In energy, inefficiency peaks at $\beta = 1$ near 4000 cost \footnote{Values for $\beta = 1$ in the energy dataset are excluded from visualizations to avoid heatmap saturation due to extremely high inefficiency values}, when adversaries exceed 85\%. In the voting dataset, inefficiency increases by 111\% in the vulnerable region, with collapse triggered beyond 70\% adversaries at $\beta \geq 0.96$. Privacy collapses earlier: $\beta \geq 0.3$ at 50\% adversaries in the high privacy-preserving signal with 50\% adversaries, and $\beta \geq 0.5$ in the low privacy-preserving signal.

Discomfort cost (Figure \ref{fig:discomfort}) shows a consistent decline as adversarial intensity increases. In resilient regions, typically below 30–50\% adversary presence and $\beta < 0.7$, discomfort initially remains high but declines gradually, then sharply in collapse regions, as adversary scale and severity increase. In the energy dataset, discomfort remains high only below 30\% adversaries and drops by 8\% before collapsing to near-zero. The voting dataset follows a similar pattern, with discomfort gradually decreasing and fully eliminated in collapse. The privacy datasets exhibit earlier collapse ($\beta > 0.3$, $>$60\% adversaries), resulting in faster discomfort decline.

Compromised discomfort cost (Figure \ref{fig:compromised}) increases with adversarial presence. In the energy dataset, it remains low for $\beta < 0.8$ even when all agents are adversarial, but rises sharply when the adversary scale exceeds 40\% at $\beta > 0.9$. In voting, the cost remain low under mild attacks ($\beta < 0.2$, adversary scales $<20\%$) and increase gradually beyond 50\% at $\beta > 0.7$. Privacy datasets show a steeper rise, peaking at 9 (high privacy) and 10 (low privacy) when 90\% of agents are adversarial. Resilience disappears at high adversary densities ($>80\%$), even under low severity, and collapse emerges at low $\beta$ when agent compromise is widespread. While energy and voting are more sensitive to $\beta$, privacy is primarily affected by adversary scale.

Interestingly, a noticeable resilience lag exists between the system-level inefficiency and the agent-level discomfort and compromise costs. In all datasets, a substantial portion of configurations classified as "Vulnerable" or even "Collapsed" by inefficiency remain in the "Resilient" state when evaluated by discomfort metrics. For instance, in the energy dataset, while only 6\% of configurations are in vulnerability based on inefficiency, 59\% fall under vulnerability based on discomfort—indicating that discomfort degrades far earlier. Similarly, in the voting dataset, 21\% of configurations are classified as vulnerable by inefficiency, compared to 70\% by discomfort. This pattern reveals a lag of up to 40–60\% in the transition from individual to system-level degradation, with discomfort and compromise metrics acting as early-warning indicators long before global inefficiency surfaces. This insight underlines the importance of incorporating agent-centric metrics for proactive resilience monitoring.

\begin{figure*}[htbp]
  \centering
  \begin{subfigure}{0.92\textwidth}
    \centering
    \includegraphics[width=\textwidth]{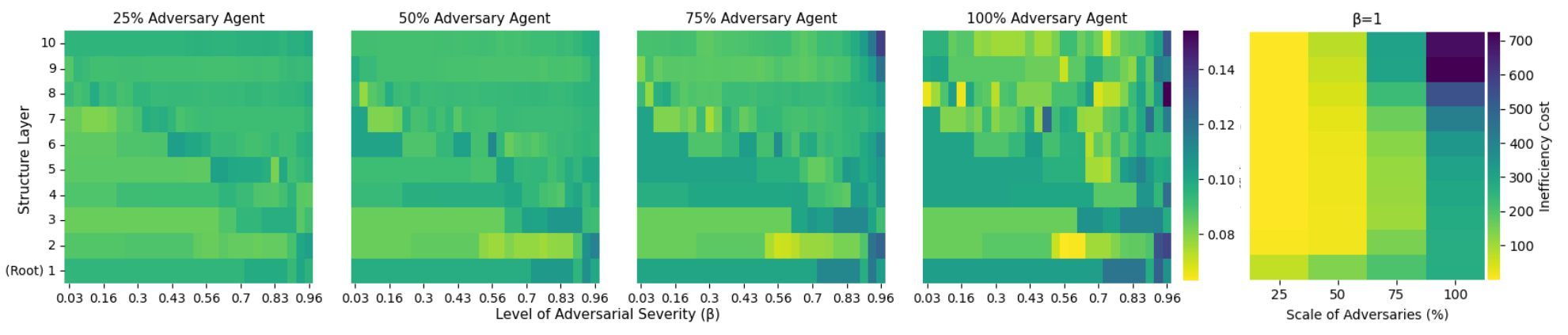}
    \caption{Energy dataset.}
    \label{layer_energy}
  \end{subfigure}

  \begin{subfigure}{0.9\textwidth} 
    \centering
    \includegraphics[width=\textwidth]{layerwise_voting.pdf}
    \caption{Voting dataset.}
    \label{layer_score}
  \end{subfigure}

  \begin{subfigure}{0.9\textwidth} 
    \centering
    \includegraphics[width=\textwidth]{layerwise_high.pdf}
    \caption{Privacy dataset (High privacy-preserving).}
    \label{layer_high}
  \end{subfigure}
  
  \begin{subfigure}{0.9\textwidth} 
    \centering
    \includegraphics[width=\textwidth]{layerwise_low.pdf}
    \caption{Privacy dataset (Low privacy-preserving).}
    \label{layer_low}
  \end{subfigure}

\caption{Inefficiency costs across hierarchical structure layers under various adversarial configurations.}
\label{fig:layers}
\end{figure*}

\subsection{Pareto Optimality Analysis}
Pareto optimality analysis identifies fronts and knee points reflecting optimal trade-offs under varying adversarial conditions. Figure \ref{fig:pareto} shows how system inefficiency relates to discomfort experienced by legitimate agents across different severity levels and adversary scales.

Figure \ref{fig:pareto_level} focuses on the impact of adversarial severity ($\beta$) on the tolerated scale of adversaries. The voting dataset maintains stable knee points across 50–60\% adversary ratios over a broad severity range ($0.1 \leq \beta \leq 0.7$), with consistent fronts even at high $\beta$. The privacy dataset shows distinct patterns: the high privacy-preserving signal tolerates 70\% adversaries at $\beta < 0.5$, decreasing to 50\% at higher severities; the low privacy-preserving signal exhibits smoother transitions with average tolerance near 68\%.

Figure \ref{fig:pareto_adv} evaluates the impact of adversarial scale on tolerated severity. For visual clarity, only 20 adversary scales are shown per dataset. In energy, knee points are stable at $0.03 < \beta < 0.3$ for up to 90\% adversaries, increase to $\beta = 0.9$ at lower scales ($<40\%$), and decline to $\beta < 0.1$ at full scale. Voting shows higher tolerance, with knee points extending $\beta = 0.8$ even under full adversaries. Privacy dataset shows greater variability: the high privacy-preserving configuration reaches $\beta = 0.2$ for 50–90\% adversaries, while the low privacy-preserving case peaks at $\beta = 0.36$ for 20\% adversaries and declines to $\beta = 0.13$ at full scale. Detailed Pareto plots are in Appendix \ref{appendix:pareto-details}.

\begin{figure*}[htbp]
  \centering
      \includegraphics[width=\textwidth]{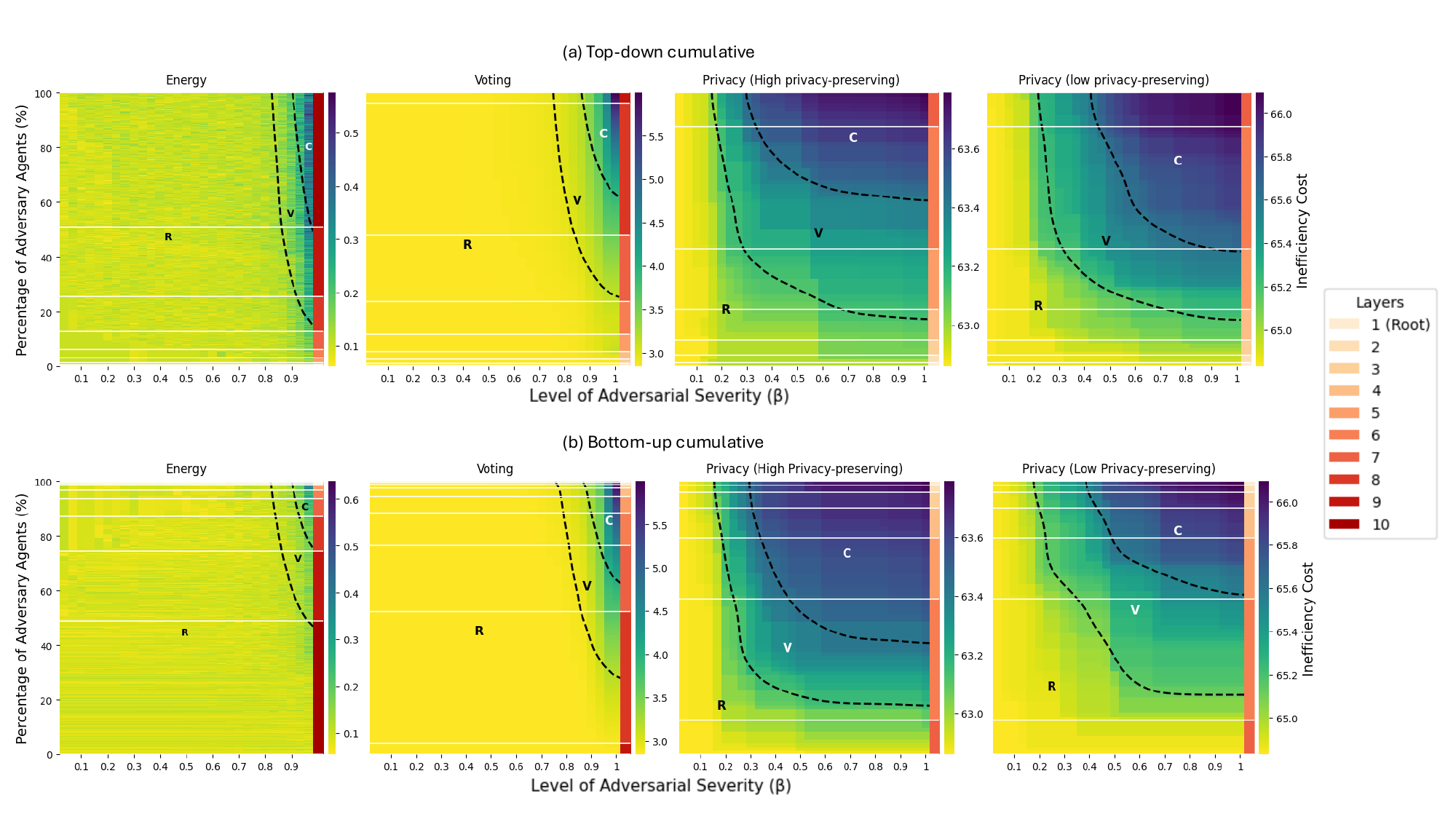}
      \caption{Inefficiency cost across hierarchical structure layers under various adversarial configurations in energy, voting and privacy datasets.}
      \label{fig:cumulative}
\end{figure*}

\subsection{Structure Analysis}
This section analyzes how hierarchical structures influence inefficiency costs across datasets using two approaches: layer-wise and cumulative structural analysis.

\subsubsection{Layer-wise Structural Analysis}
Figure \ref{fig:layers} presents the inefficiency costs across hierarchical layers under adversary scales of 25\%, 50\%, 75\%, and 100\%. For layers with few agents, intermediate adversary ratios are approximated by averaging to the closest feasible configurations.

In the energy dataset (Figure \ref{layer_energy}), with 1,000 agents over 10 layers, inefficiency remains low (below 0.16) across moderate adversarial scales and gradually increases with severity ($\beta < 1.0$). The minimum inefficiency occurs at layer 2 under 100\% adversaries and $\beta=0.56$, while inefficiency peaks at layer 8 under $\beta=0.96$, marking a 144\% increase from the minimum. Profiles remain smooth under 25\% and 50\% adversary scales but fluctuate more at 75\% and 100\%.  Under extreme severity ($\beta=1$), inefficiency surges sharply, reaching over 723 at full adversarial saturation. With this severity, the root layer shows moderate inefficiency, exceeding layers 2–7 at 75\% and matching layers 2 and 3 at 100\%, followed by a steady increase down the hierarchy. Despite layer 10 hosting the largest agent population, its inefficiency values are slightly lower than those of layer 9.

In the voting dataset (Figure \ref{layer_score}), with 266 agents over 9 layers, inefficiency remains stable across all layers under low adversarial severity ($\beta \leq 0.3$). As severity increases, costs escalate, particularly at the root (Layer 1) and Layers 2 and 3, which consistently experience higher inefficiency than deeper layers. The maximum inefficiency  is observed at layer 8 under 100\% adversaries, where agent density is highest. Layer 9,despite its depth, shows lower inefficiency due to a smaller agent count. At 25\% and 50\% adversary scales, inefficiency profiles remain relatively smooth; however, at 75\% and 100\%, cost escalations become pronounced, particularly in upper and middle layers.

In the privacy dataset (Figures \ref{layer_high} and \ref{layer_low}), with 72 agents across 7 layers, inefficiency remains low and uniform under low severities. In the high privacy-preserving configuration, the root layer consistently incurs higher costs than subsequent layers across all adversary scales, with inefficiency peaking at layer 6 (the layer with the largest number of agents) under $\beta \geq 0.7$. Layer 7, although deeper, has lower costs due to fewer agents. In the low privacy-preserving configuration, similar trends are observed with occasional fluctuations between layers 5 and 6 at higher adversary scales. Overall, inefficiency profiles remain structurally consistent between both privacy settings, though severity levels accelerate cost increases in the high-privacy case.

\subsubsection{Cumulative Structural Analysis}
Cumulative analysis evaluates how adversarial influence propagates through hierarchical structures in two configurations: top-down (root-to-leaf) and bottom-up (leaf-to-root). Figure \ref{fig:cumulative} shows inefficiency across layers under both directions.

In the top-down positioning, the energy dataset remains resilient up to layer 7 with 20\% adversaries. Collapse occurs at layer 10 when the adversary ratio exceeds 50\% and $\beta > 0.9$. In voting, resilience holds through layer 4, with vulnerability at layer 5 and collapse at layers 6–7 under high severity ($\beta > 0.9$). In the privacy datasets, both signals remain resilient in the top four layers, with vulnerability emerging at layer 4 for $\beta > 0.7$. Collapse appears at layer 6 in the high privacy-preserving signal with over 60\% adversaries, and slightly earlier in the low privacy-preserving signal with 50\% adversaries.

In the bottom-up direction, the energy dataset shows a narrower vulnerability and collapse region. Vulnerability begins at layer 9 with 50\% adversaries, and collapse follows at layer 10 under high severity ($\beta > 0.9$). Voting results are similar across both directions, with transitions driven more by adversary ratio than structural depth. Vulnerability emerges at layer 6 with 20\% adversaries, and collapse follows at layer 5 with 60\%. In privacy, vulnerability appears at layer 6 in both signals; collapse occurs at layer 6 (high privacy-preserving) and layer 5 (low privacy-preserving).

\begin{figure*}[tb]
  \centering
  \begin{subfigure}{0.98\textwidth} 
    \includegraphics[width=\textwidth]{gaussian_inefficency_nobeta1.pdf}
    \centering
    \caption{Inefficiency cost}
    \label{fig:gaussian_inefficiency}
  \end{subfigure}

  \begin{subfigure}{0.98\textwidth}
    \includegraphics[width=\textwidth]{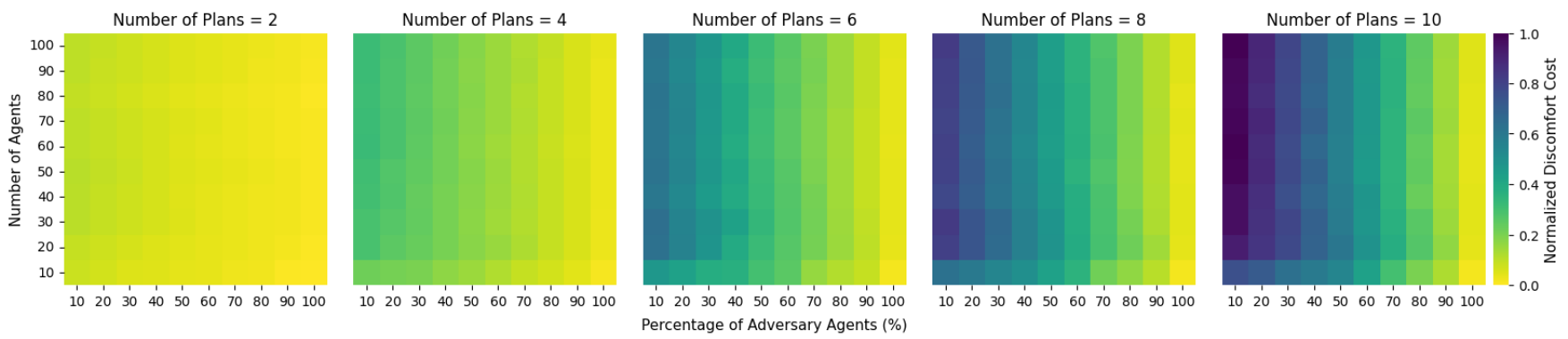}
    \centering
    \caption{Discomfort cost}
    \label{fig:gaussian_discomfort}
  \end{subfigure}

\caption{Normalized inefficiency and discomfort costs on synthetic Gaussian data across varying numbers of agents, plan options per agent, and adversary ratios.}

\label{fig:gaussian}
\end{figure*}

\subsection{Scalability Analysis with Synthetic Data}
To benchmark scalability and validate trends from real datasets, we use a synthetic Gaussian dataset to systematically vary agent population (10–100) and plan count (2–10) under different adversarial ratios. This controlled setup reveals optimization performance across configurations under adversarial pressure\footnote{Results exclude $\beta=1$ to avoid saturation effects and highlight subtler trends.}.

Figure \ref{fig:gaussian_inefficiency} shows normalized inefficiency costs across agent counts, adversary ratios, and plan numbers. Inefficiency increases notably beyond 60\% adversarial presence and is most pronounced in systems with smaller agent populations. Configurations with fewer plans exhibit high inefficiency even under moderate adversarial influence, indicating lower tolerance. At 100\% adversaries, inefficiency peaks in 10-agent systems. In contrast, larger agent populations and broader plan spaces significantly enhance resilience, suppressing inefficiency across most adversarial scales.

Figure \ref{fig:gaussian_discomfort} presents normalized discomfort costs under the same conditions. Discomfort increases with the number of plans. Systems with two plans show a gradual decline in discomfort as adversary ratios increase, while those with more plans experience sharper reductions. Discomfort variation expands with plan complexity, indicating greater sensitivity. The number of agents has minimal impact, though the 10-agent setting occasionally shows marginally lower discomfort than larger systems.

\subsection{Summary of Findings and Discussion} 
The key findings of this work can be summarized as follows: 
\begin{enumerate}
\item The interplay of adversarial scale and severity determines the resilience, vulnerability and collapse of distributed multi-objective optimization, which is strongly influenced by the optimization scenario.

\item Distributed multi-objective optimization can predominantly remain resilient, even for high adversarial scales or severity.

\item Adversarial attacks trigger high comfort losses by legitimate agents as collective compromises that reduce the likelihood of collapse for higher vulnerability and resilience. 

\item Comfort compromises of legitimate agents for preserving system efficiency under adversarial attacks are predominantly required for high adversarial severity.

\item Pareto optimal points for adversarial severity levels and adversarial scales are mainly in the resilience trajectory. However, Pareto optimal points for high adversarial scales can expand to vulnerability and collapse trajectories. 

\item High adversarial scales reduce the comfort compromises required by the legitimate agents in the Pareto optimal points for adversarial severity levels that can be tolerated. 

\item High adversarial severity levels reduce the system efficiency in the Pareto optimal points for adversarial scales that can be tolerated.

\item Lower hierarchical levels with higher scales of agents within hierarchical structures of distributed multi-objective optimization are more vulnerable to adversarial attacks than top levels with lower scales of agents.

\item A top-down positioning of adversary agents within hierarchical structures of distributed multi-objective optimization is more impactful on system performance: higher vulnerability, likelihood of collapse, and lower resilience.

\item Systems with a small number of agents and low plan diversity are more susceptible to inefficiency increases under adversarial pressure, even at moderate adversary ratios.

\item Broader plan spaces and larger agent populations enhance system resilience in distributed multi-objective optimization, significantly suppressing inefficiency across adversarial configurations.

\end{enumerate}
The findings of this paper can be used to develop and enhance corrective self-healing strategies~\cite{nejad2021enhancing} that are cost-effective in practice. They can also be used to design incentive mechanisms \cite{bhattacharya2022incentive} that ensure agents comply to certain standards of safety in critical infrastructures such as smart grids. For instance, one challenge of fault-correction mechanism is the timely detection of adversary agents to mitigate their impact~\cite{nejad2021enhancing}. Apparently, redundancy mechanisms and rollback operations orchestrated by monitoring mechanisms are resource-intensive~\cite{nejad2021enhancing, liang2023self}. They require frequent checks that involve computations and exchange of messages and they usually rely on static thresholds or even manual operations~\cite{liang2023self}. This is where the insights of this work can find applicability: these mechanisms can adapt based on the status of the system, for instance, whether it is in the resilience, vulnerability and collapse state. Knowing apriori that an optimization process can tolerate certain adversarial scales and severity can simplify and reduce the costs of applying prevention and mitigation measures. It can also provide new insights for security policies, for instance, prioritizing the protection of agents at the top of the hierarchical optimization structure with stronger security safeguards and resources allocated for this purpose~\cite{enoch2022integrated}. While creating these strategies does not fall into the scope of this paper, it is part of the future work to pursue.

\section{Conclusion and Future work} \label{sec6}
This study provides a comprehensive analysis of resilience, vulnerability, and collapse dynamics in multi-agent distributed optimization under adversarial conditions. By systematically examining adversary scale, severity, and network structure, we identify critical thresholds where systems transition from stability to failure. These findings offer actionable guidance for designing and enhancing the performance of recovery and healing strategies.
A key contribution of this work is the release of a large-scale benchmark dataset, generated from over 112 million experiments using the proposed adversarial model. This dataset supports systematic evaluation of adversarial impacts and facilitates reproducible research across domains of distributed optimization.

Although the adversarial model is designed to be general-purpose, the evaluation can be extended to other algorithms in future work. Additionally, the current experiments model adversarial behavior with static severity levels, which may not fully capture dynamic or strategic adversary actions. Furthermore, resilience has been analyzed under hierarchical network structures, leaving the behavior under alternative topologies an open area for exploration.

Future work will focus on embedding adaptive monitoring and mitigation mechanisms into real-time distributed systems. Investigating more complex network structures, dynamic adversarial strategies, and diverse application domains will further advance the development of robust, fault-tolerant optimization systems capable of maintaining performance under adversarial conditions.

\section*{Acknowledgments}
This project is funded by a UKRI Future Leaders Fellowship (MR-/W009560-/1): `\emph{Digitally Assisted Collective Governance of Smart City Commons--ARTIO}'.
The authors would like to thank Thomas Wellings and Chuhao Qin for their feedback on the paper, Abhinav Sharma for technical support in experimentation and Srijoni Majumdar for support with the voting dataset.

\bibliographystyle{IEEEtran}
\bibliography{references}

\vspace{0.5cm}

\begin{center}
   \appendix 
\end{center}
\vspace{0.5cm}
\section*{Appendix A: Cost Function Definitions}
\label{appendix:cost-functions}

This appendix presents the mathematical definitions of the cost functions used in the optimization experiments. The functions are inherited from the original I-EPOS framework \cite{pournaras2018decentralized} and are instantiated to reflect the characteristics of each application domain.

\subsection*{Variance}
The variance cost measures the dispersion of the aggregated global response \(\GlobalResponse \in \mathbb{R}^d\) and is used in the energy and Synthetic Gaussian datasets. It is computed as:
\[
f_{\text{var}} = \frac{1}{d} \sum_{j=1}^{d} \left( \GlobalResponse_j - \bar{\GlobalResponse} \right)^2
\]
where:
\begin{itemize}
    \item \( \GlobalResponse_j \) is the aggregated global response at dimension \( j \),
    \item \( \bar{\GlobalResponse} \) is the mean of the global response across all dimensions,
    \item \( d \) is the dimensionality of the response.
\end{itemize}

\subsection*{Residual Sum of Squares (RSS)}
The RSS cost quantifies the squared difference between the scaled global response \(\GlobalResponse \in \mathbb{R}^d\) and a predefined system-wide target signal \( T \in \mathbb{R}^d \). It is used in the voting and privacy datasets and is defined as:
\[
f_{\text{RSS}} = \left( s(\GlobalResponse) - s(T) \right)^T \left( s(\GlobalResponse) - s(T) \right)
\]
where \( s(\cdot) \) denotes the scaling function applied to both vectors to improve shape alignment.
\vspace{0.5cm}

\section*{Appendix B: Pareto Optimality Visualizations} \label{appendix:pareto-details}
This appendix presents detailed visualizations of Pareto fronts for 10 selected adversarial severity levels (\(\beta\)) and 10 adversary population scales across the energy, voting, and privacy datasets. Each subfigure illustrates the trade-off between inefficiency cost and the discomfort cost of legitimate agents. Red lines indicate the non-dominated Pareto fronts, while red boxes mark the knee points, identified using the Minimum Manhattan Distance (MMD) method. These visualizations complement the analysis in Section~\ref{sec5}, offering deeper insights into system behavior under varying adversarial conditions.
\vspace{0.5cm}
\section*{Appendix C: Pareto Optimality of Total Agents}\label{appendix:pareto-details2}
Figure~\ref{fig:pareto_total} shows the trade‐off between system inefficiency and total‐agent discomfort, with Pareto knee points identified for both varying severity (a) and adversary scale (b).
\vspace{0.5cm}

\begin{figure}[h]
  \centering
  \begin{subfigure}{0.49\textwidth}
    \includegraphics[width=\linewidth]{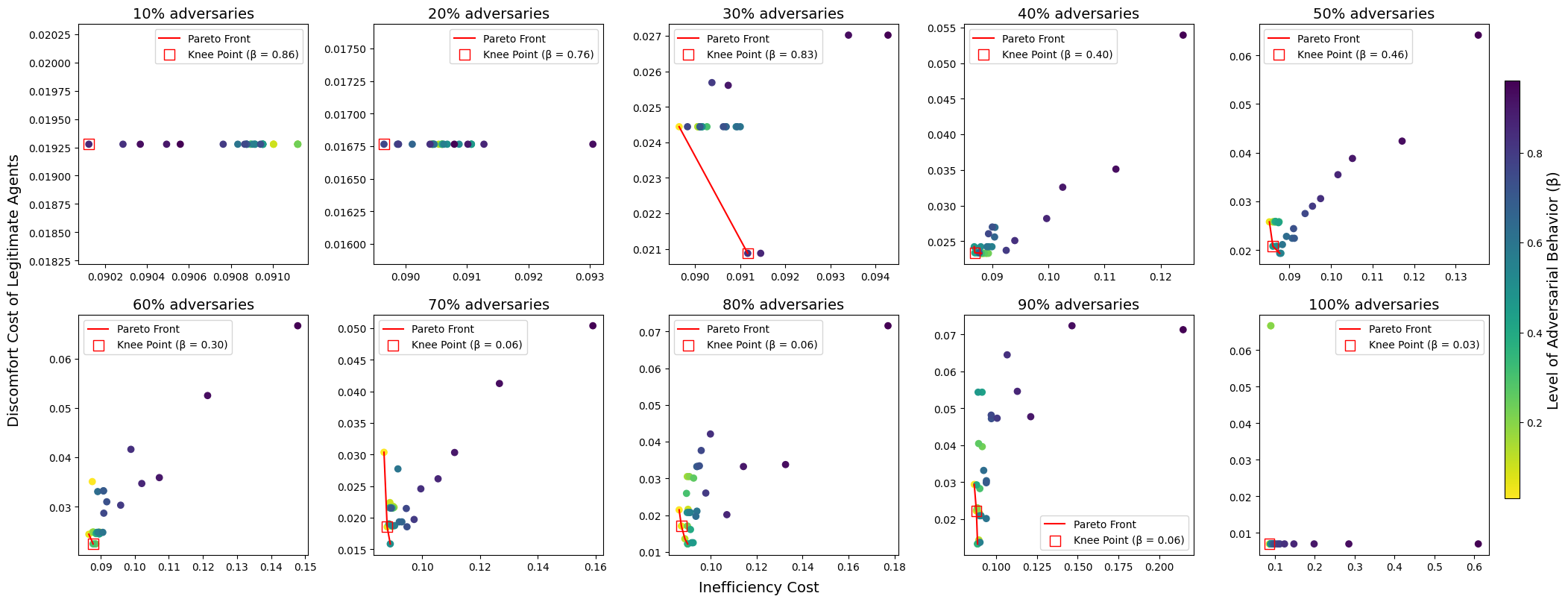}
    \caption{Energy dataset — Pareto fronts and knee points across adversary population scales (10\% to 100\%).}
    \label{fig:energy_adv_appendix}
  \end{subfigure}
  \vfill
  \begin{subfigure}{0.49\textwidth}
    \includegraphics[width=\linewidth]{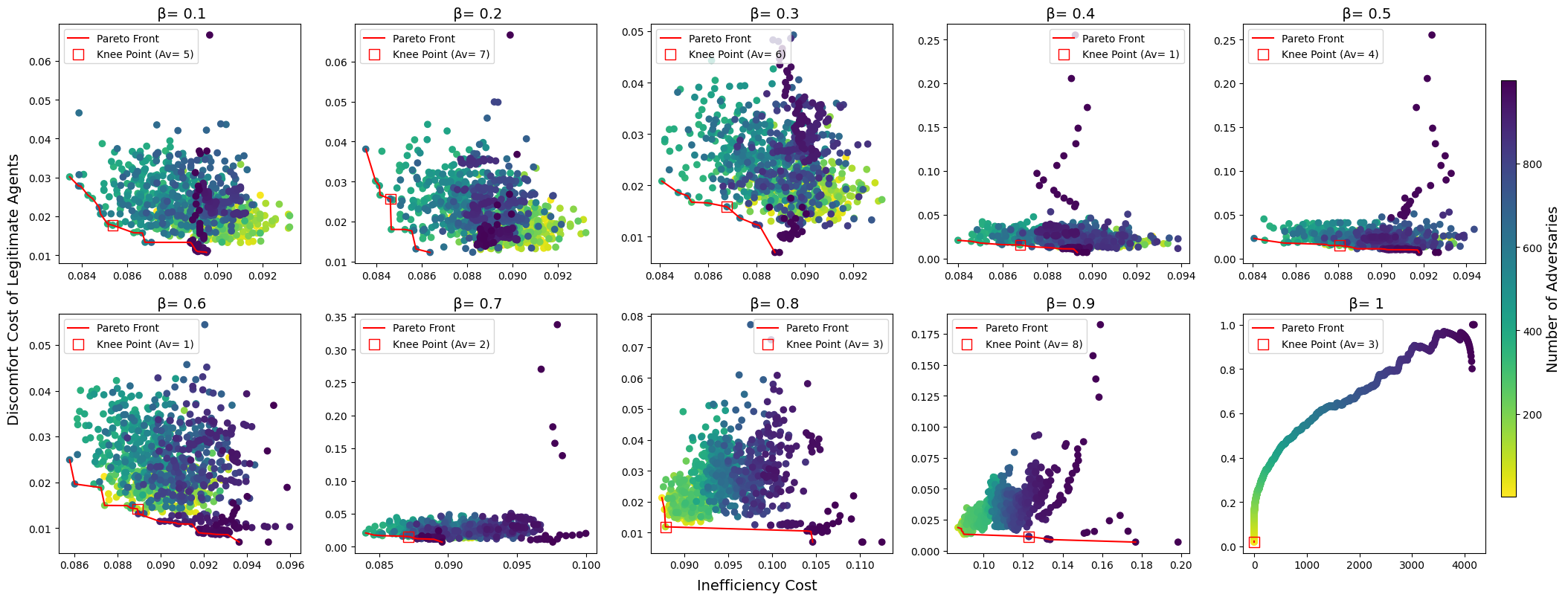}
    \caption{Energy dataset — Pareto fronts and knee points across adversarial severity levels (\(\beta = 0.1\) to 1.0).}
    \label{fig:energy_beta_appendix}
  \end{subfigure}
  \caption{Pareto optimality analysis for the Energy dataset.}
\end{figure}

\begin{figure}[h]
  \centering
  \begin{subfigure}{0.49\textwidth}
    \includegraphics[width=\linewidth]{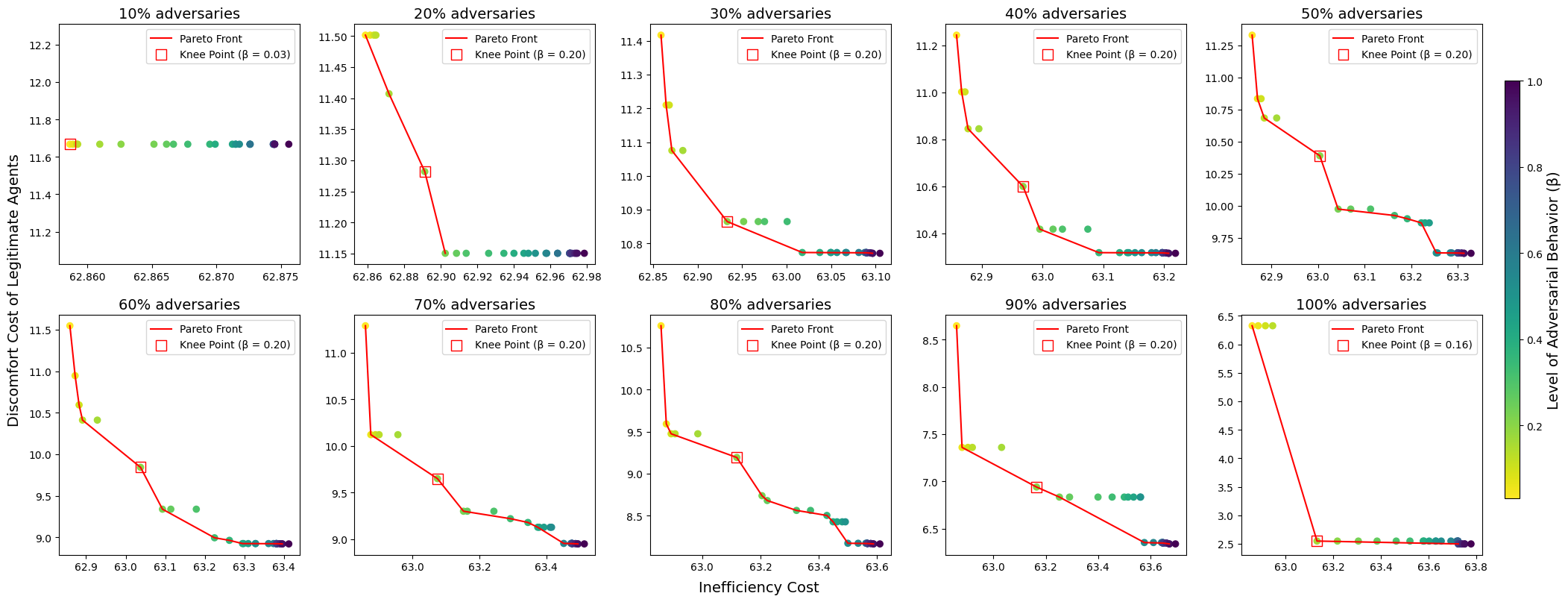}
    \caption{Privacy dataset (high signal) — Pareto fronts and knee points across adversary population scales (10\% to 100\%).}
    \label{fig:privacy_high_adv}
  \end{subfigure}
  \vfill
  \begin{subfigure}{0.49\textwidth}
    \includegraphics[width=\linewidth]{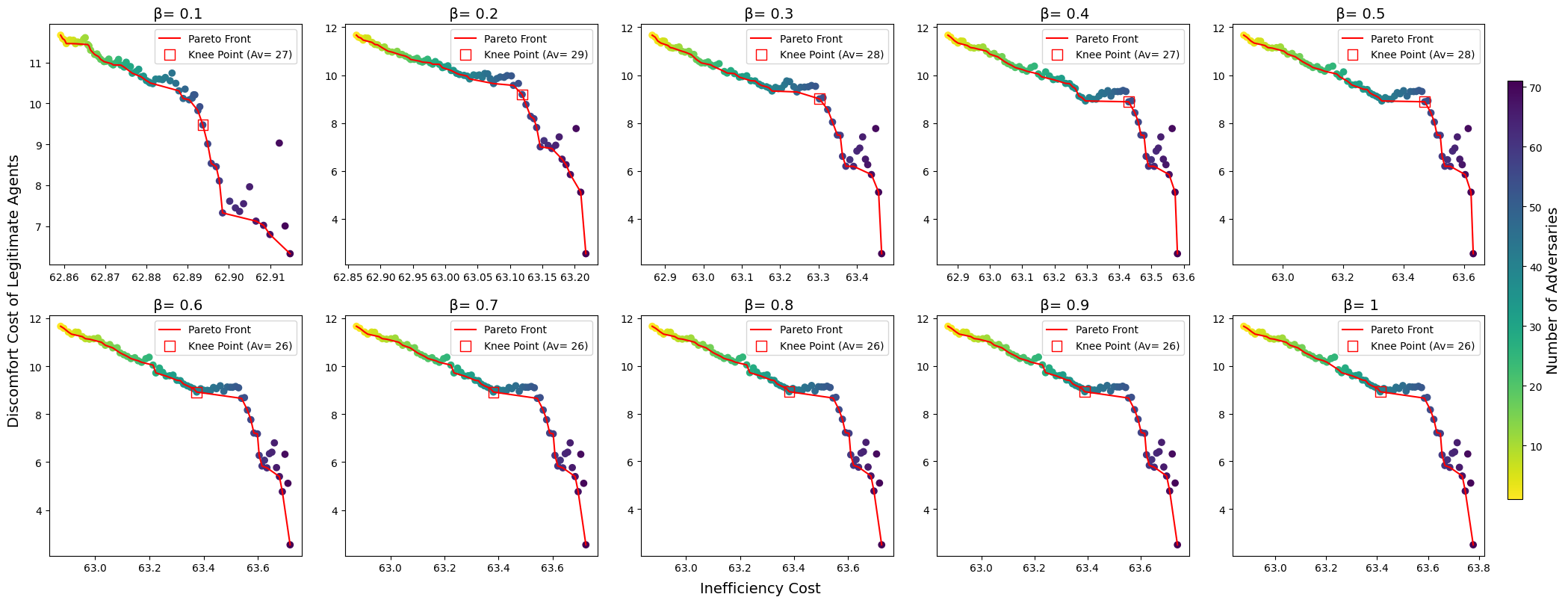}
    \caption{Privacy dataset (high signal) — Pareto fronts and knee points across adversarial severity levels (\(\beta = 0.1\) to 1.0).}
    \label{fig:privacy_high_beta}
  \end{subfigure}
  \caption{Pareto optimality analysis for the high privacy-preserving signal in the privacy dataset.}
\end{figure}

\begin{figure*}[htbp]
  \centering
  \begin{subfigure}{0.49\textwidth}
    \includegraphics[width=\linewidth]{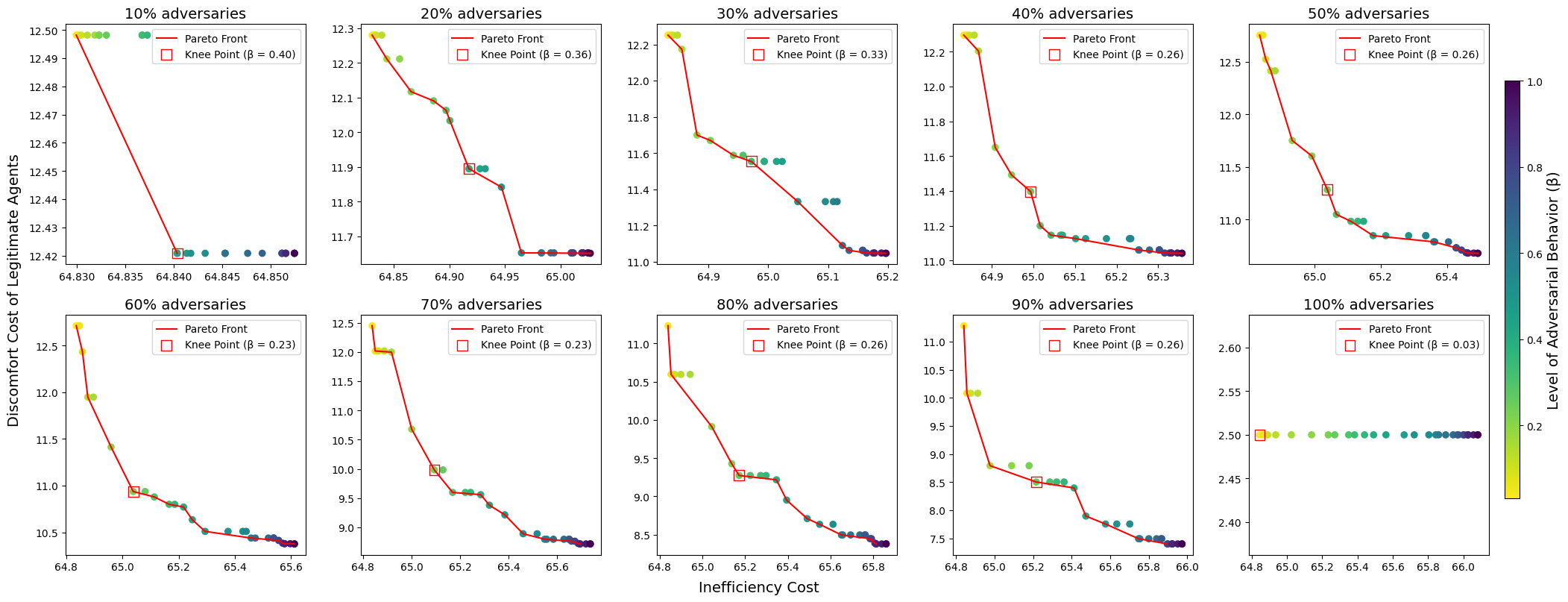}
    \caption{Privacy dataset (low signal) — Pareto fronts and knee points across adversary population scales (10\% to 100\%).}
    \label{fig:privacy_low_adv}
  \end{subfigure}
  \hfill
  \begin{subfigure}{0.49\textwidth}
    \includegraphics[width=\linewidth]{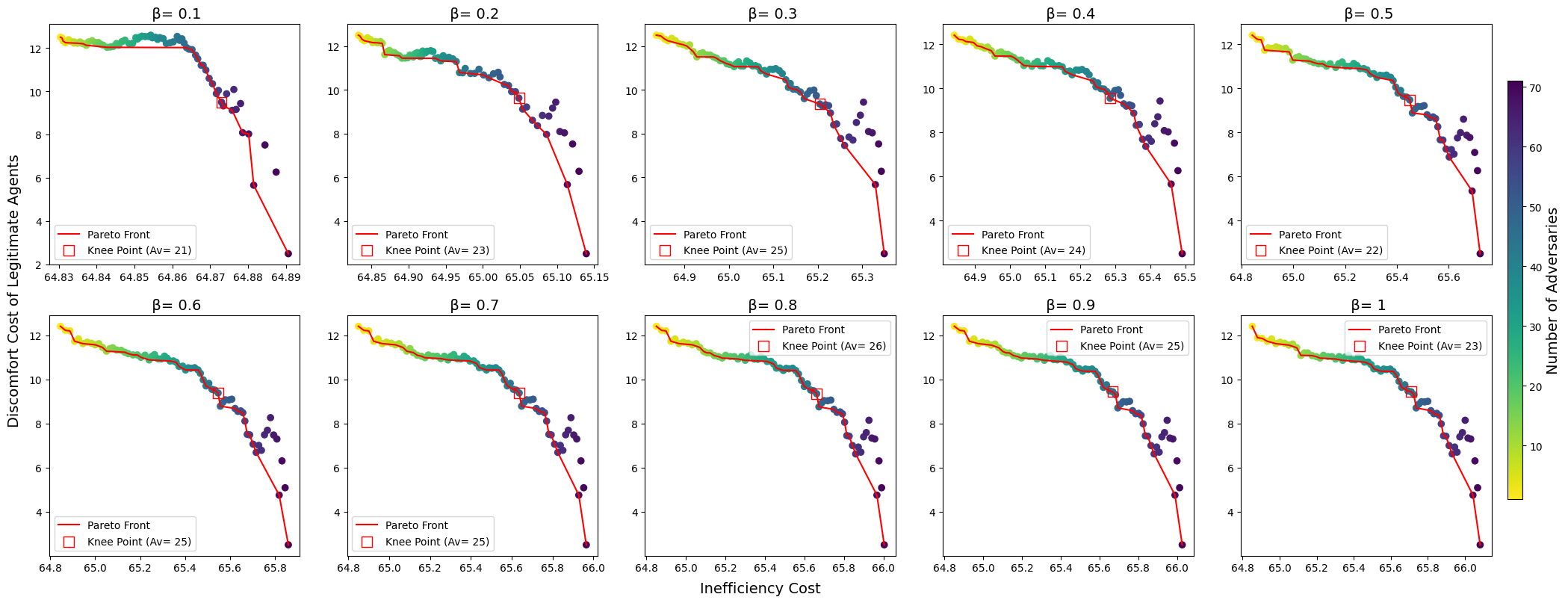}
    \caption{Privacy dataset (low signal) — Pareto fronts and knee points across adversarial severity levels (\(\beta = 0.1\) to 1.0).}
    \label{fig:privacy_low_beta}
  \end{subfigure}
  \caption{Pareto optimality analysis for the low privacy-preserving signal in the privacy dataset.}
\end{figure*}

\begin{figure*}[htbp]
  \centering
  \begin{subfigure}{0.49\textwidth}
    \includegraphics[width=\linewidth]{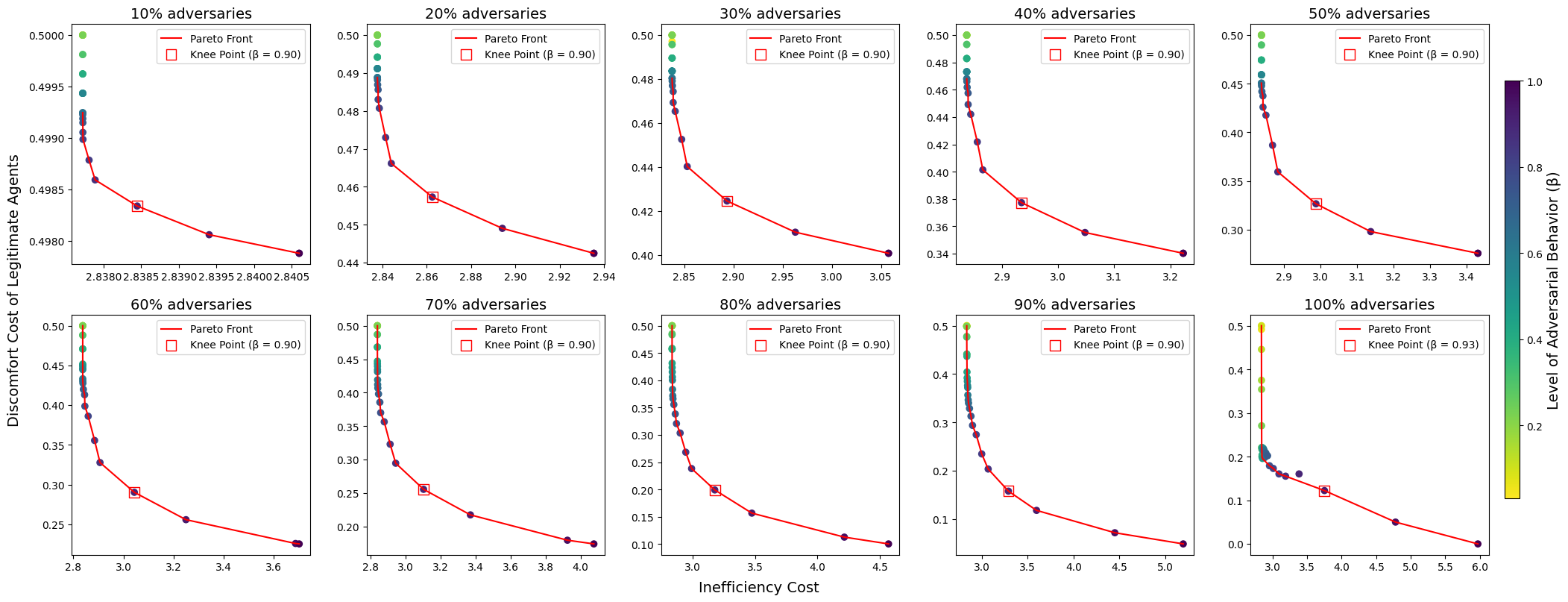}
    \caption{Voting dataset — Pareto fronts and knee points across adversary population scales (10\% to 100\%).}
    \label{fig:voting_adv_appendix}
  \end{subfigure}
  \hfill
  \begin{subfigure}{0.49\textwidth}
    \includegraphics[width=\linewidth]{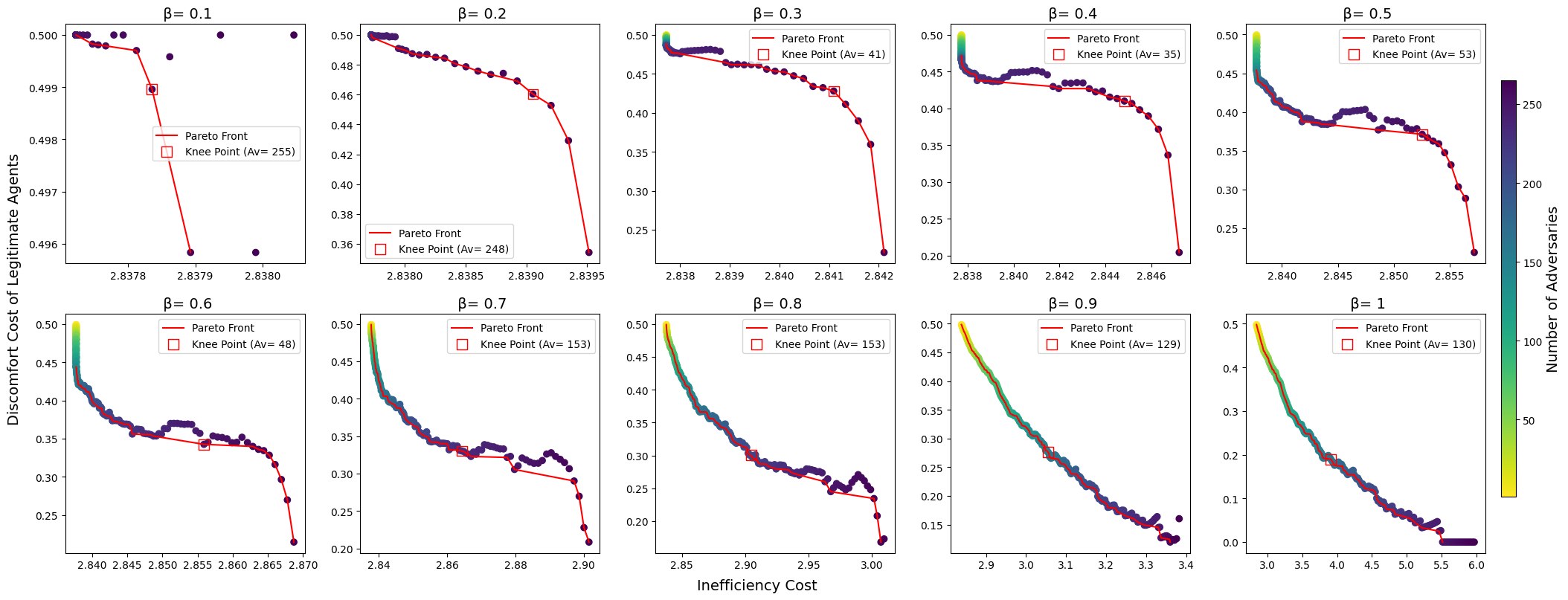}
    \caption{Voting dataset — Pareto fronts and knee points across adversarial severity levels (\(\beta = 0.1\) to 1.0).}
    \label{fig:voting_beta_appendix}
  \end{subfigure}
  \caption{Pareto optimality analysis for the voting dataset.}
\end{figure*}

 \begin{figure*}[b]
  \centering
  \begin{subfigure}{0.98\textwidth} 
    \includegraphics[width=\textwidth]{pareto1_total.pdf}
    \centering
    \caption{Pareto front and knee points across adversarial severity.}
    \label{fig:total_pareto1}
  \end{subfigure}

  \begin{subfigure}{0.98\textwidth}
    \includegraphics[width=\textwidth]{pareto2_total.pdf}
    \centering
    \caption{Pareto front and knee points across scales of adversaries.}
    \label{fig:total_pareto2}
  \end{subfigure}

\caption{The Pareto Optimality of The energy voting, and privacy datasets}
\label{fig:pareto_total}
\end{figure*}

\end{document}